\documentclass[reprint,amsmath,amssymb,aps]{revtex4-1}

\usepackage{color}
\usepackage{amsmath}
\usepackage{graphicx}
\usepackage{dcolumn}
\usepackage{bm}
\usepackage{booktabs}
\usepackage{multirow} 

\begin{document}
\preprint{APS/123-QED}

\title{Quantum Topological Boundary States in Quasi-crystal}

\author{Yao Wang$^{1,2}$, Yong-Heng Lu$^{1,2}$, Jun Gao$^{1,2}$, Ke Sun$^{1,2}$, Zhi-Qiang Jiao$^{1,2}$, Hao Tang$^{1,2}$}
\author{Xian-Min Jin$^{1,2}$}
\email{xianmin.jin@sjtu.edu.cn}

\affiliation{
	$^1$State Key Laboratory of Advanced Optical Communication Systems and Networks, School of Physics and Astronomy, Shanghai Jiao Tong University, Shanghai 200240, China\\
	$^2$Synergetic Innovation Center of Quantum Information and Quantum Physics, University of Science and Technology of China, Hefei, Anhui 230026, China}

\date{\today}

\maketitle

\textbf{Topological phase, a novel and fundamental role in matter, displays an extraordinary robustness to smooth changes in material parameters or disorder. A crossover between topological physics and quantum information may lead to inherent fault-tolerant quantum simulations and quantum computing. Quantum features may be preserved by being encoded among topological structures of physical evolution systems. This requires us to stimulate, manipulate and observe topological phenomena at single quantum particle level, which, however, hasn't been realized yet. Here, we address such a question whether the quantum features of single photons can be preserved in topological structures. We experimentally observe the boundary states of single photons and demonstrate the performance of topological phase on protecting the quantum features in quasi-periodic systems. Our work confirms the compatibility between macroscopic topological states and microscopic single photons on a photonic chip. We believe the emerging `quantum topological photonics' will add entirely new and versatile capacities into quantum technologies.}\\

\noindent The study of topological phenomenon in quantum systems, deriving from the quantum Hall effect, has revealed the fundamental role of topological phases in matter \cite{review1,review2,review3,Hall1}. The hallmark of the novel phases in topological insulator is the emergence of the topological protected state on the system edge or surface comparing to the ordinary insulator. In the view of band structures, the localized edge or surface states appear at the interface between two topologically distinct systems closing to energy gaps, such as the chiral modes of the integer quantum Hall effect \cite{Hall1} and the Dirac cone in topological insulator \cite{Graphene,3D}.

The topological phase displays an extraordinary robustness to smooth changes in material parameters or disorder and cannot be changed unless the system undergoes a topological transition. Therefore the topological phase holds the promise of inherently fault-tolerant quantum simulations and computing \cite{computing}. The topological quantum computation, using non-Abelian anyons to store and manipulate quantum information in a nonlocal manner, is becoming a promising candidate towards quantum computers in theoretical predictions, but is still in its infancy in experimental implementations \cite{Top_Q_com_1,Top_Q_com_2,Top_Q_com_book}. Alternatively, another potential approach is to directly protect quantum features in topological structures of physical evolution systems.

In contrast to the extremely demanding experimental environment in condensed-matter or high-energy physics, light, as one specific spectrum range of electromagnetic waves, is easy to be created and detected, and has been employed to investigate quantum topological phenomena, for example, the Hall effect \cite{Topo_hall_1,Topo_intro_1,Topo_intro_2,Topo_intro_3}, edge states \cite{Topo_edge_1,Topo_edge_2,Topo_edge_3,Topo_edge_4,Topo_edge_5}, topological insulators \cite{Topo_TIth_1,Topo_TIth_2,Topo_TIth_3,Topo_TIex_1,Topo_TIex_3} and the Weyl points \cite{Topo_weyl_1,Topo_weyl_2}. Meanwhile, single photon, as single particle of light, is a promising platform for quantum simulation and quantum computing \cite{photon_sim}. Thus, photon can be simultaneously compatible to both topological physics and quantum information, which suggests quantum topological photonics as a new frontier for exploring both fundamental problems and potential applications.

\begin{figure*}
	\centering
	\includegraphics[width=1.8\columnwidth]{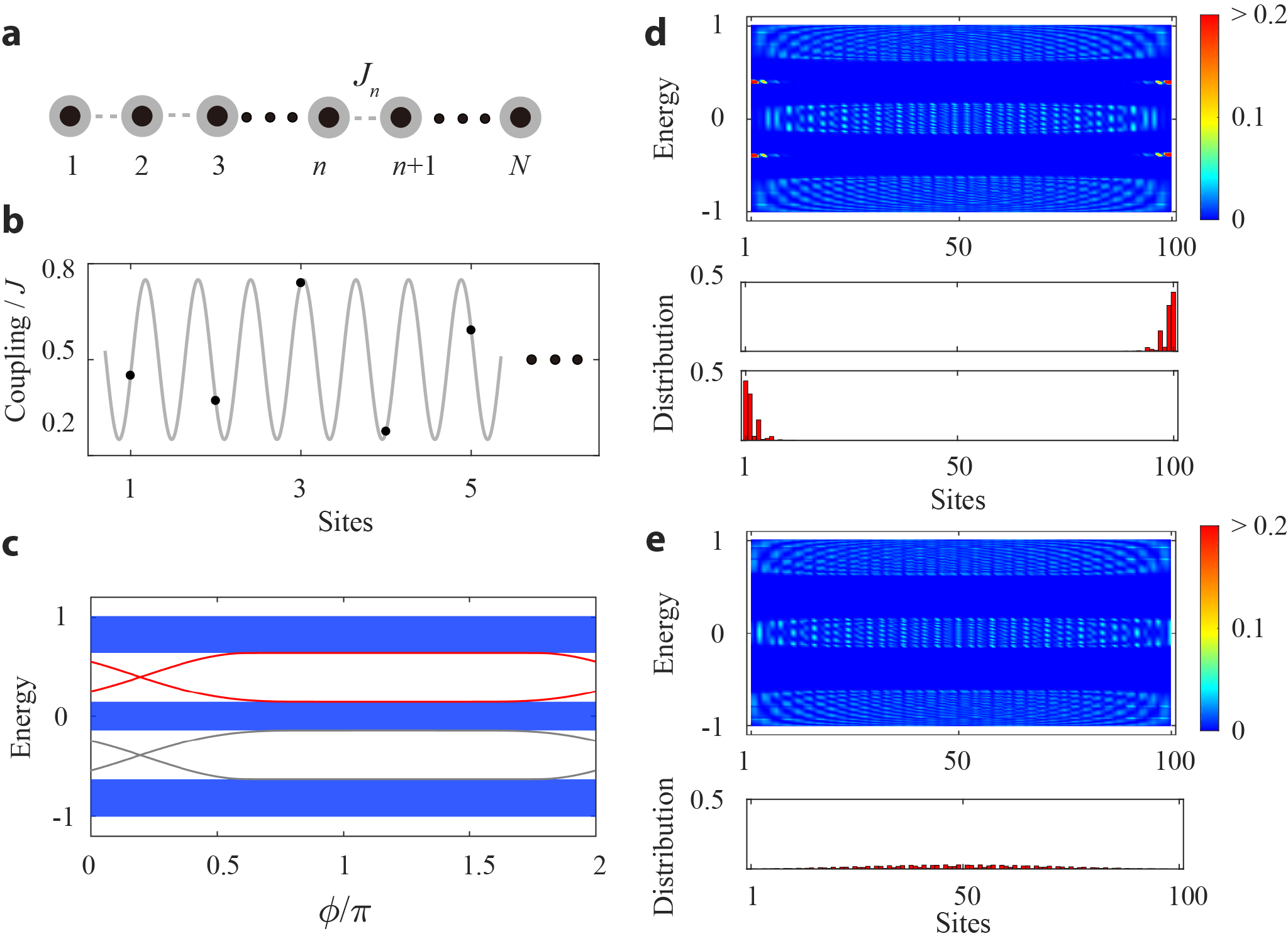}\\
	\caption{\textbf{Band structure of quasi-crystal and the local density of states.} \textbf{a-b,} The coupling strength of a one-dimension quasi-crystal lattice is modulated in the form of a cosine incommensurate. \textbf{c,} The Floquet band structure of quasi-crystal as a function of $\phi $ for $t=0.5$, $\lambda=0.5$, $b=(\sqrt{5}+1)/2$(the gold mean). The gaps between bands are crossed by two nontrivial topological edge modes. \textbf{d,} The spatial distribution of the eigenstates for the specific structure ($\phi =0.2\pi$) as a function of energy. Two types of boundary models are particularly prominent among all, in which single photons are confined on the boundary. \textbf{e,} The spatial distribution of the eigenstates for $\phi =0.9\pi$. There is no boundary state such that the single photons are scattered all over the lattice for all models.}
	\label{f1}
\end{figure*}

In this work, we experimentally observe the topological boundary states of single photons, and investigate the quantum features in a quasi-periodic lattice of up to 100 sites on a photonic chip. We show that the site number is an extra decisive element, besides the quasi-periodicity modulation parameter, for the topological phase. Adding or removing one site at the end of system can transform the band structure and change the boundary state type. Besides, we observe that the quantum features of single photon can be well preserved in quasi-crystal system via boundary states. 

The quasi-periodic system, ordered but not periodic, was found to offer a convenient platform for the study of topological phases \cite{QC1}. And the quasi-crystal can be characterized by the topological invariant that is usually attributed to systems of a dimension higher than their physical dimension \cite{QC2}. With the quasi-periodicity of Harper model (also known as Aubry-Andr{\'e} model) \cite{AA1,AA2}, the quasi-crystal possesses robust boundary states and the topological characteristics is similar to the two-dimension integer quantum Hall effect \cite{QC1,QC2}. 

As is shown in Fig.\ref{f1}(a), our work is based on the `off-diagonal' version of Harper model, which could be described by the Hamiltonian
\begin{equation}
    \label{eq1}
     H=\sum_{n}J(n)a_na_{n+1}^++h.c.,
\end{equation}
where $a^+_n$($a_n$) is the creation(annihilation) operator at site $n$, $J$ presents the coupling strength between the adjacent sites, defined by $J(n)=t(1+\lambda cos(2\pi bn+\phi))$. In this expression, $t$ is the average coupling strength, $\lambda $ presents the modulation amplitude, and $b$ is the parameter controlling the order of crystal.

\begin{figure*}
	\centering
	\includegraphics[width=1.5\columnwidth]{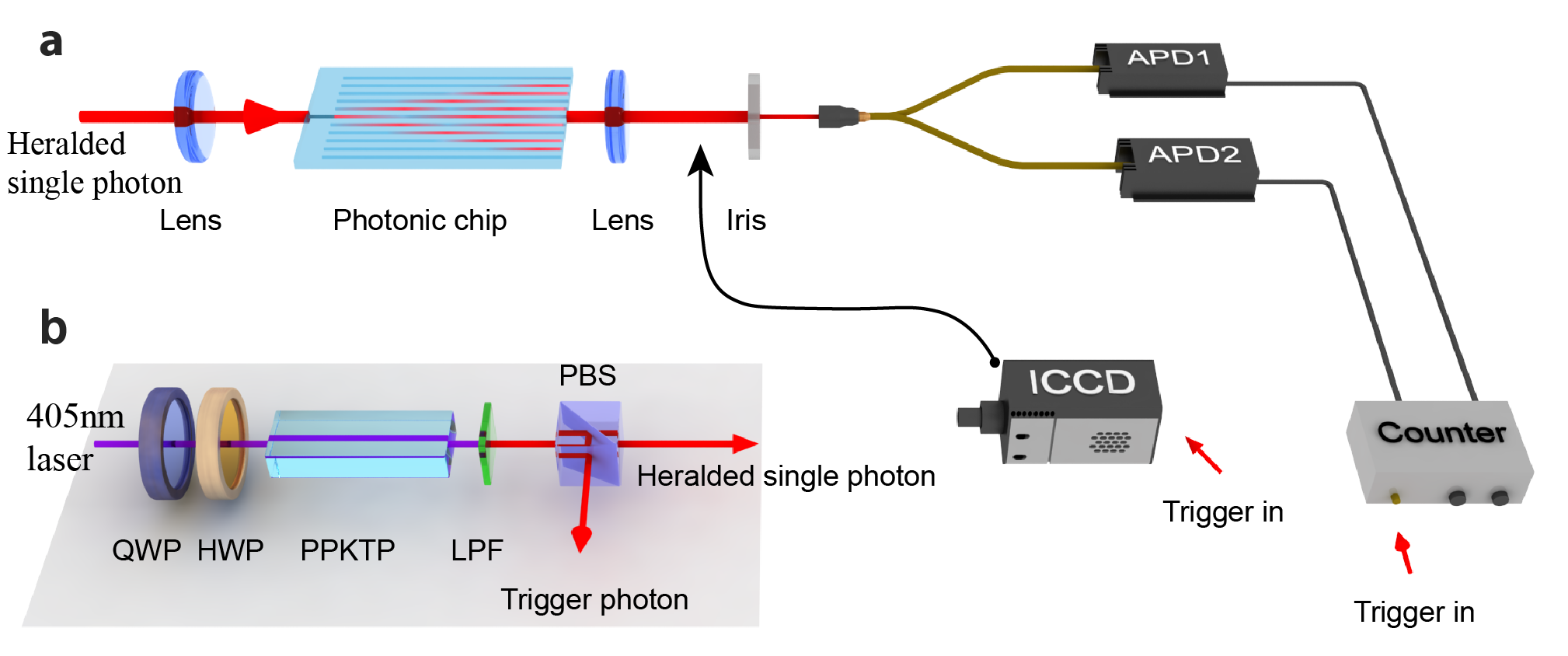}\\
	\caption{\textbf{Experimental setup.
		}\textbf{a,} Different photonic quasi-crystals are fabricated in the same photonic chip. Heralded single photons are injected into the lattices after be focused by a lens, and are collected at the output facet behind an iris which ensures only the single photons out of the chosen site are collected. The second-order anti-correlation parameter $\alpha$ of single photons is measured with APDs using the Hanbury-Brown-Twiss interference. An ICCD is used to observe single-photon distribution probability before the iris. \textbf{b,} The generation of the heralded single photon and the trigger photon.}
	\label{f2}
\end{figure*}

We set the parameter $b$ to be irrational. As the result, the coupling strength $J$ is quasi-periodic varying with the site $n$ as shown in Fig.\ref{f1}(b), and the system is non-periodic but has long-range order. The system exhibits trivial and nontrivial topological properties with different additional degrees of freedom, behaving as the modulation phase. The topological properties contribute to a higher dimension via tunning the modulation phase $\phi$.

\begin{figure*}
	\centering
	\includegraphics[width=1.9\columnwidth]{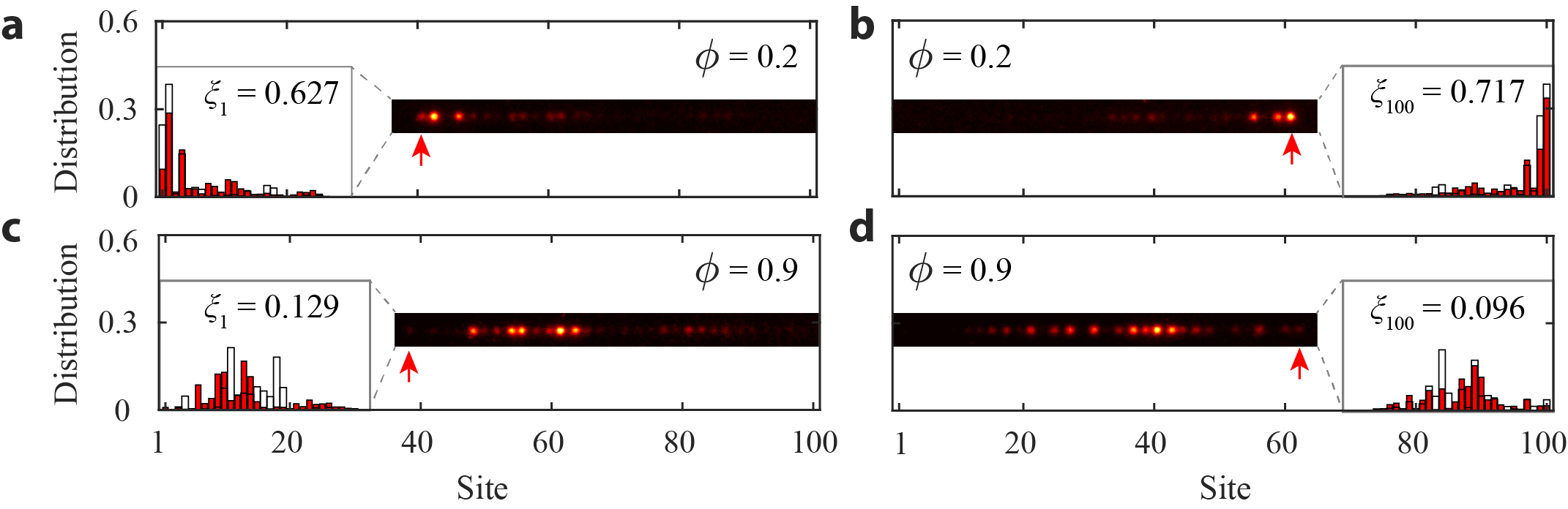}\\
	\caption{\textbf{Experimental results of single-photon distribution probability.} \textbf{a-b,} Two types of topological boundary states are observed when we inject heralded single photons into the lattice with parameter $\phi=0.2\pi$. \textbf{c-d,} The heralded single photons are no longer confined to the boundary of the system with parameter $\phi=0.9\pi$. The red arrows point out the input sites. The red histogram represents the experimental results and the white histogram is our simulated results for reference.}
	\label{f3}
\end{figure*}

\begin{figure*}
	\centering
	\includegraphics[width=1.9\columnwidth]{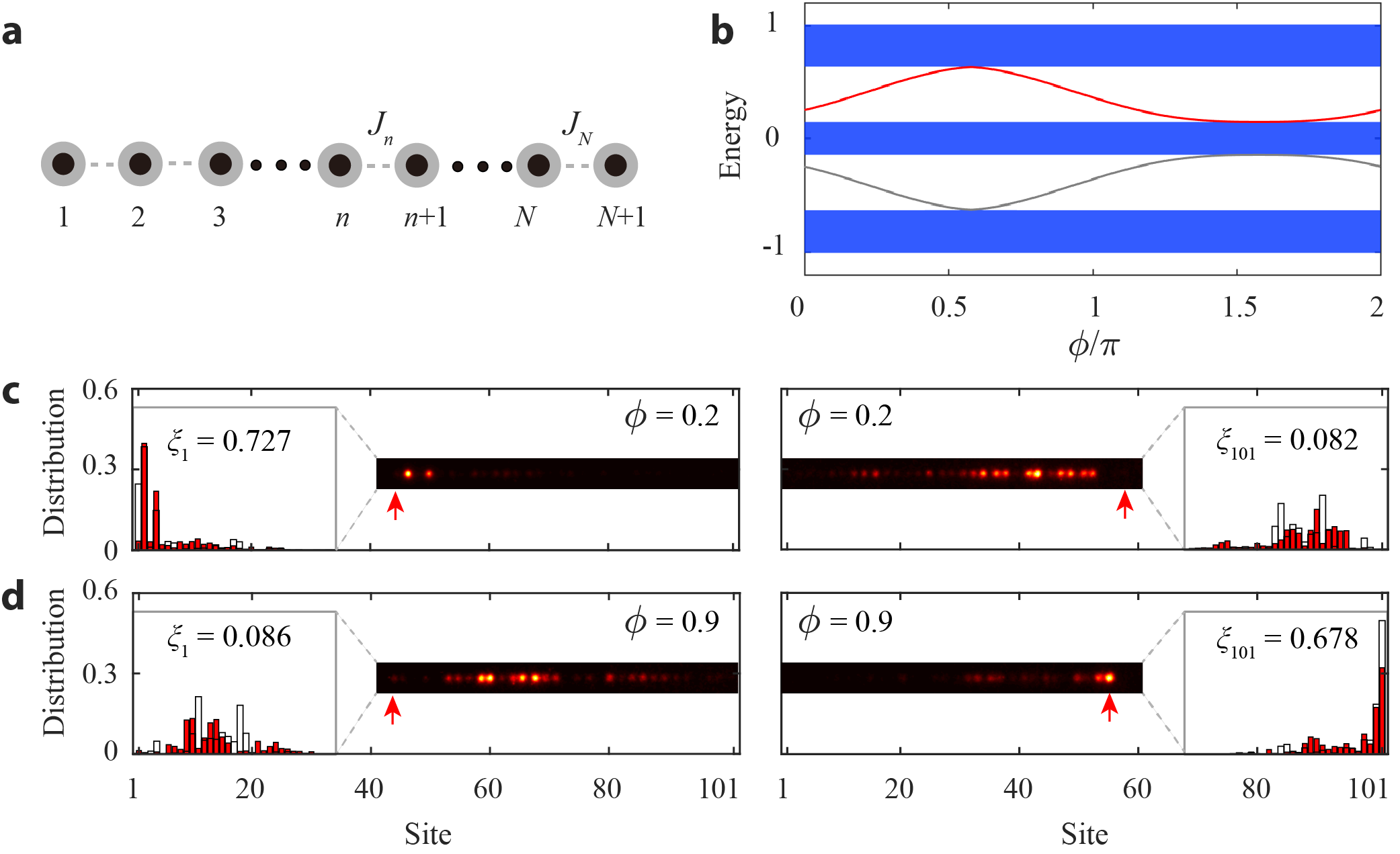}\\
	\caption{\textbf{Site-number dependence of quasi-crystal band structure.} Adding one more site to the end of the lattice (\textbf{a}), the band structure transforms to a new form (\textbf{b}). Comparing to the 100-sited lattice, either one or zero type of boundary state exists. Only the left boundary state appears with the modulation phase $\phi=0.2\pi$ (\textbf{c}), and only the right boundary state can be observed with the modulation phase $\phi=0.9\pi$ (\textbf{d}). The red arrows point out the input sites. The red histogram represents the experimental results and the white one is our simulated results for reference.}
	\label{f4}
\end{figure*}

We show the calculated Floquet band structure of the photonic quasi-crystal as a function of $\phi$ in Fig.\ref{f1}(c). The band structure shows that the gaps between bands are crossed by two nontrivial topological boundary modes, in which the states are localized on the boundary of the system. To reveal the detailed message in the band structure for a given $\phi$, we show the spatial distribution of the eigenstates as a function of energy by the local density of states for $\phi =0.2\pi$ (see Fig.\ref{f1}(d)), which is defined by $D_n(E)=\sum_{m}\delta(E-E_m)|\varphi_n^{(m)}|^2$ ($E_m$ is the energy of the $m$th eigenstates and $\varphi_n^{(m)}$ is its wave function \cite{QC2}). Four boundary states among all modes are particularly prominent. Single photons can be confined on the edge of structure under the norm of the boundary state. As a comparison, we plot the local density of states for $\phi =0.9\pi$ in Fig.\ref{f1}(e), in which there is no boundary state between the bands. Single photons distribute over all the sites for each eigenstate.

In our experiment, we fabricate the quasi-crystals in borosilicate glass using femtosecond laser direct writing technique (see Methods) \cite{fabri_1,fabri_2,fabri_3,fabri_4}. We implement two 100-sited photonic lattices with modulation phases $\phi=0.2\pi$ and $0.9\pi$. Heralded single photons (see Methods) are injected into the leftmost site (n=1) and the rightmost site (n=100) respectively for both lattices, and the outgoing probability distribution is measured after the single photon has hopped for 35 mm. The single-photon outgoing distribution is measured with ICCD, and switchable photon statistics is obtained by APDs and an FPGA counter, see Methods and Fig.\ref{f2} for more details.

The experimental results of single-photon distribution probability are shown in Fig.\ref{f3}. From the Floquet band structure shown in Fig.\ref{f1}(a), two different types of boundary states both exist in the system if the modulation phase $\phi=0.2\pi$. When the single photon is injected into the lattices from the leftmost site (n=1), as shown in Fig.\ref{f3}(a), the single photons occupy several sites with high probability at the left boundary at the outgoing facet. The right topological boundary state [see Fig.\ref{f3}(b)] can also be observed when the single photon is initially injected into the rightmost site (n=100). The system becomes topological trivial when we set the modulation phase $\phi=0.9\pi$. Unlike the single-photon behavior in topological nontrivial phase, the photons are no longer confined to the boundary of the system [see Figs.\ref{f3}(c)-\ref{f3}(d)], regardless whether the system is excited at the leftmost site (n=1) or the rightmost site (n=100).

To quantify the localization of the outgoing single photons for the boundary states, we calculate the generalized return probability $\xi_j$ defined by $\xi_j=\sum_{j-d}^{j+d}I_i/\sum_{1}^{n}I_i$, which quantifies the probability of the single photon that remains within a small distance $d$ from the injection site $j$, and the $d=7$ is the width of the localized boundary state \cite{QC2}. In our experiment, $\xi_1=0.627$ and $\xi_{100}=0.717$ for the quasi-crystal with $\phi=0.2\pi$. As a comparison, $\xi_1=0.129$ and $\xi_{100}=0.096$ for the system is in topological trivial phase.

\begin{table}
	\centering
	\par
	\caption{\textbf{The measured $\alpha$ for the single photons going out from boundary states.} $B^{\phi}_{N_{L/R}}$ represents the topological boundary states of single photons with different structure parameters. $\phi$ indicates the modulation phase. $N$ is the site number of the lattice, and the $L$ ($R$) denotes the left (right) boundary state.}
	
	\begin{tabular}{p{1.3cm}<{\centering} p{1.3cm}<{\centering} p{1.3cm}<{\centering} p{1.3cm}<{\centering} p{1.3cm}<{\centering} p{1.3cm}<{\centering}}
		\hline\noalign{\smallskip}
		\hline\noalign{\smallskip}
		
		$B^{\phi }_{N_{L/R}}$ & $input$ & $B^{0.2}_{100_L}$ & $B^{0.2}_{100_R}$ & $B^{0.2}_{101_L}$ & $B^{0.9}_{101_R}$ \\[0.4cm]
		
		$\alpha$ & 0.0166    & 0.0143     & 0.0093     &  0.0210    &  0.0152    \\[0.15cm]
		& (0.0048)  & (0.0027)  & (0.0020)  &  (0.0074)  &  (0.0041) \\[0.15cm]
		
		\hline\noalign{\smallskip}
		\hline\noalign{\smallskip}
	\end{tabular}
	
	\label{tab1}
\end{table}

We can see that the quasi-crystal can be modified in topological trivial and nontrivial phase by tuning the modulation phase $\phi$. There are two other parameters also contributing to the form of the band structure, the modulation amplitude $\lambda$ and the site number $N$. The former controls the width of the gap between bands. The latter can lead to the transformation of the boundary states in the band structure. If we add one more site to the end of the lattice [see Fig.\ref{f4}(a)], the band structure transforms to a new form, as is shown in Fig.\ref{f4}(b). One more site renders that there is only one topological nontrivial model crossing the bands. The state of this model is localized in the left boundary, but not in the right boundary, when $\phi=0.2\pi$. The right boundary state appears when $\phi=0.9\pi$.

We fabricate two 101-sited photonic lattices with the modulation phase $\phi=0.2\pi$ and $0.9\pi$ respectively to experimentally demonstrate the predicted transformation of the band structure.
The experimental single-photon distribution probabilities are depicted in Fig.\ref{f4}(c)-\ref{f4}(d). The result is consistent with the prediction resulting from the band structure, the distribution only remained tightly localized at either left boundary for $\phi=0.2\pi$ with $\xi_1=0.727$ and $\xi_{101}=0.082$ or right boundary for $\phi=0.9\pi$ with $\xi_1=0.086$ and $\xi_{101}=0.678$. This is a clear signature of the existence of only one boundary state for certain modulation phase $\phi$.

To explore whether the topological nontrivial boundary state can preserve the single-photon feature after the single photon outgoing from the lattices, we quantify the single-photon feature with the second-order anti-correlation parameter $\alpha$. And the parameter $\alpha$ tends to be zero for ideal single photon \cite{alpha}. 

By using Hanbury-Brown-Twiss interferometer, we are able to measure the second-order anti-correlation parameter $\alpha=p_1p_{123}/p_{12}p_{13}$ with avalanche photodiode (APD) 1, 2 and 3, where APD3 is the detector for the trigger photon (not shown in Fig.\ref{f2}). In our experiment, we measure $\alpha$ for the sites that have the highest distribution probability. As is shown in Table.\ref{tab1}, the measured $\alpha$ are found being small and having no distinct changes before and after the photonic chip, which confirms that the single-photon feature can be well preserved.

In conclusion, we demonstrate the appearance of topological boundary states of single photons in quasi-crystal and the performance of the topological phase on protecting quantum features. By using single-photon imaging technique and Hanbury-Brown-Twiss interferometer, we are able to experimentally confirm the compatibility between macroscopic topological states and microscopic single photons on a photonic chip. Our work represents an attempt to directly protect quantum features in quantum topological boundary states, combining topological physics and quantum information. The emerging quantum topological photonics opens up an immediate demand to fundamentally investigate quantum topological boundary states in multi-photon regime, and to explore potential applications in quantum simulation and quantum computing in both theory and experiment.

\subsection*{Acknowledgments}
The authors thank Jian-Wei Pan for helpful discussions. This research is supported by the National Key Research and Development Program of China (2017YFA0303700), National Natural Science Foundation of China (Grant No. 61734005, 11761141014, 11690033, 11374211), the Innovation Program of Shanghai Municipal Education Commission, Shanghai Science and Technology Development Funds, and the open fund from HPCL (No. 201511-01), X.-M.J. acknowledges support from the National Young 1000 Talents Plan.\\

\subsection*{Methods}
{\bf Fabrication and measurement of the quasi-crystal on a photonic chip:} We design the quasi-crystal lattice structure according to the characterized relationship between the coupling coefficients and the separation of adjacent waveguides. The lattices are written into borosilicate glass substrate (refractive index $n_0=1.514$) with femtosecond laser (power 10W, wavelength 1026nm, SHG wavelength 513nm, pulse duration 290fs, repetition rate 1MHz). We reshape the focal volume of the beam with a cylindrical lens, and then focus the beam inside the borosilicate substrate with a 50X objective lens (NA=0.55),  A high-precision three-axis translation stage is in charge of moving the photonic chip during fabrication with a constant velocity of 10mm/s.

In the experiment, we inject the heralded single photons into the rightmost or leftmost sites in the photonic chip using a 20X objective lens. After a total propagation distance of 35 mm through the lattice structures,  the outgoing probability distributions of heralded single photons are observed using a 10X microscope objective lens and the ICCD camera.\\

{\bf The generation and imaging of the heralded single photon:} The single-photon source with the wavelength of 810nm is generated from periodically-poled KTP (PPKTP) crystal via spontaneous parametric down conversion. The generated photon pairs are separated to two components, horizontal and vertical polarization, after a long-pass filter and a polarized beam splitter (PBS). One should notice that the measured patterns would come from the thermal-state light rather than single-photons if we inject only one polarized photon into the lattices without external trigger. Therefore, we inject the horizontally polarized photon into the lattices, while the vertically polarized photon acts as the trigger for heralding the horizontally polarized photons out from the lattices with a time slot of 10 ns. We capture each evolution result using the ICCD camera after accumulating in the external mode for 2000 s. The second-order anti-correlation parameter $\alpha$ is measured using APDs and an FPGA counter after accumulating for 300s.

\end{document}